\documentclass[a4paper,aps,preprintnumbers,showpacs,twocolumn,superscriptaddress,nofootinbib,amsmath,amssymb]{revtex4-1}
\usepackage[dvips]{graphics}
\def\imo{i}

\def\Order#1{{\cal O}\left(#1\right)}

\newcommand{\rem}[1]{}
\usepackage[utf8]{inputenc}
\usepackage[T1]{fontenc}
\usepackage{cmap}
\begin{document}
\title{Further clarification on quasinormal modes/circular null geodesics correspondence}

\author{R. A. Konoplya}
%\email{roman.konoplya@gmail.com}
%\email{olexandr.zhydenko@ufabc.edu.br}
\affiliation{Research Centre for Theoretical Physics and Astrophysics, Institute of Physics, Silesian University in Opava, Bezručovo náměstí 13, CZ-74601 Opava, Czech Republic}

%\date{\today}

\begin{abstract}
The well-known duality between quasinormal modes of any stationary, spherically symmetric and asymptotically flat or de Sitter black hole
and parameters of the circular null geodesic was initially claimed for gravitational and test field perturbations. According to this duality the real and imaginary parts of the $\ell \gg n$ quasinormal mode (where $\ell$ and $n$ are multipole and overtone numbers respectively) are multiples of the frequency and instability timescale of the circular null geodesics respectively. Later it was shown that the duality is guaranteed only for test fields and may be broken for gravitational and other non-minimally coupled fields. Here, we farther specify the duality and prove that even when the duality is guaranteed it {\it may not represent the full spectrum of the $\ell \gg n$ modes}, missing the quasinormal frequencies which cannot be found by the standard WKB method. In particular we show that this always happens for an arbitrary asymptotically de Sitter black holes and further argue that, in general, this might be related to sensitivity of the quasinormal spectrum to geometry deformations near the boundaries.
\end{abstract}

%\pacs{04.30.Nk,04.50.+h}
\maketitle

%\section{Introduction}

Observations of black holes via gravitational waves and shadows \cite{Abbott:2016blz,TheLIGOScientific:2016src,EventHorizonTelescope:2019dse,Goddi:2016jrs,Bambi:2015kza} makes it interesting to understand relations between characteristics of proper frequencies of black holes, called {\it quasinormal modes} \cite{QNMreviews}, and geodesics describing propagation of light in the black hole environment. An important observation, in this context, was made in \cite{Cardoso:2008bp}, where it was stated that parameters of the unstable circular null geodesics around a stationary, spherically symmetric and asymptotically flat or de Sitter black holes, are dual to quasinormal modes  that the  black hole emits in the
$\ell \gg n$ regime, where $\ell$ is the multipole number and $n$ is the overtone. Thus, it was shown that the quasinormal frequencies of the four and higher dimensional Schwarzschild-(de Sitter) black hole are
\begin{equation}\label{QNM}
\omega_n=\Omega_c\,\ell-i(n+1/2)\,|\lambda|, \quad \ell \gg n
\end{equation}
where $\Omega_c$ is the angular velocity at the unstable null geodesics and $\lambda$ is the Lyapunov exponent. Then, in \cite{Cardoso:2008bp} it was deduced that the relation (\ref{QNM}) must be valid for all stationary, spherically symmetric asymptotically flat or de Sitter black holes.
This finding brought enormous attention and was used in a great number of consequent papers. Being unable to mention about 500 publications where this correspondence was discussed, we will review here only several most relevant contributions. Further development of the correspondence to black hole shadows and other configurations can be found in \cite{Chen:2022ynz,Li:2021zct,Chen:2021gwy,Jusufi:2020dhz,Jusufi:2019ltj,Momennia:2019cfd,Toshmatov:2019gxg,Breton:2016mqh}.

The case of axially symmetric black holes was discussed in  \cite{Cardoso:2008bp} in the  slow rotation regime, and further it was noticed that for Kerr black holes the correspondence between photon spheres and eikonal quaisnormal modes is more complicated  for an arbitrary angular momentum \cite{Yang:2012he}.  Later it was observed that the link between the null geodesics and quasinormal modes is stipulated by the history of development of black hole models rather than a strict and general constraining duality \cite{Khanna:2016yow}.
In \cite{Konoplya:2017wot} it was showed that there are counterexamples to this correspondence, related to the unusual regime of $\ell \rightarrow \infty$, which occurs, for example in theories with higher curvature corrections, such as Einstein-Gauss-Bonnet, Einstein-dilaton-Gauss-Bonnet and Einstein-Lovelock ones \cite{Konoplya:2017wot,Konoplya:2019hml}. In \cite{Konoplya:2017wot} it was concluded that the correspondence is {\it guaranteed for test fields} around stationary spherically symmetric black holes, but may be invalid for gravitational and other non-minimally coupled fields, if the effective potential is not positive definite single maximum barrier which monotonically decays at the event horizon and infinity (or de Sitter horizon). Thus, as a counterexample, the $\ell \gg 1$ potential barrier is accompanied by appearance of the negative gap near the event horizon which becomes infinite leading to instability at some values of the Gauss-Bonnet/Lovelock coupling constants \cite{Dotti:2004sh,Takahashi:2010gz,Konoplya:2020bxa,Konoplya:2017lhs}. The quasinormal modes of gravitational perturbations in this regime evidently cannot be described by ``the well-behaved'' effective potential.

We believe that owing to great interest to this duality and frequent usage of it in both directions (from quasinormal modes to null geodesics and vice versa), further specification of the cases for which the correspondence works and understanding whether it is strict and constraining is worth of consideration. Here we will consider situation when the effective potential has the usual single maximum and monotonic decay near the boundaries, consequently, the geodesics are in agreement with the eikonal quasinormal frequencies. Nevertheless, we will show that there are $\ell \gg n$ modes which cannot be reproduced via this duality and they dominate at late times. In other words, while the formula (\ref{QNM}) is still valid, it does not represent the full spectrum in the regime of high multipole numbers, that is, $n$ is not representing the overtone number in eq. (\ref{QNM}) anymore.

%``''

%\section{WKB arguments}

A static, spherically symmetric metric in $D$ -dimensional spacetime has the form:
\begin{equation}\label{metric}
d s^2 = f(r) d t^2 - \frac{1}{g(r)} d r^2 -r^2 d\Omega_{n}^2,
\end{equation}
where  $f(r)$ and $g(r)$ represent solutions of the gravitational and matter field equations under consideration and $d\Omega_n^2$ is a $(n=D-2)$-dimensional sphere.

Following Cardoso et. al. \cite{Cardoso:2008bp}, one can see that the principal Lyapunov exponent \cite{Lyapunov} for null geodesics around a static,  spherically symmetric metric (\ref{metric}) is
\begin{equation}\label{GenLyap}
\lambda = \frac{1}{\sqrt{2}}\sqrt{-\frac{r_c^2}{f_c}\left(\frac{d^2}{dr_*^2}\frac{f}{r^2}\right)_{r=r_c}},
\end{equation}
where the tortoise coordinate is defined as
%\begin{equation}\label{tort}
$dr/dr_*=\sqrt{g(r)f(r)}$.
%\end{equation}
The coordinate angular velocity for the null geodesics is
\begin{equation}\label{angularvel}
\Omega_c = \frac{f_c^{1/2}}{r_c},
\end{equation}
where $r_{c}$ is the radius of the circular null geodesics,  satisfying the equation
\begin{equation}\label{circulareq}
 2f_c=r_cf'_c.
\end{equation}

After separation of variables the gravitational or matter fields perturbation equations can usually be reduced to the following wave-like form
\begin{equation}
\left(\frac{d^2}{d r_*^2} +\omega^2 -V(r_*)\right)\Psi(r_*)=0,
\end{equation}
where $\Psi$ is the wave functions.

The arguments in \cite{Cardoso:2008bp}  are based on the similarity between the WKB treatment suggested in \cite{Schutz:1985zz} (and further developed in \cite{Iyer:1986np,Kokkotas:1988fm,Konoplya:2003ii,Matyjasek:2017psv,Konoplya:2019hlu}) for finding quasinormal modes
and the above equations of the null geodesics.
However the WKB formula is developed for effective potentials which have the form of the potential barrier with a single extremum outside the event horizon and approach constant values at the event horizon and spacial infinity or de Sitter horizon for asymptotically de Sitter spacetimes.

\begin{figure}
\resizebox{\linewidth}{!}{\includegraphics{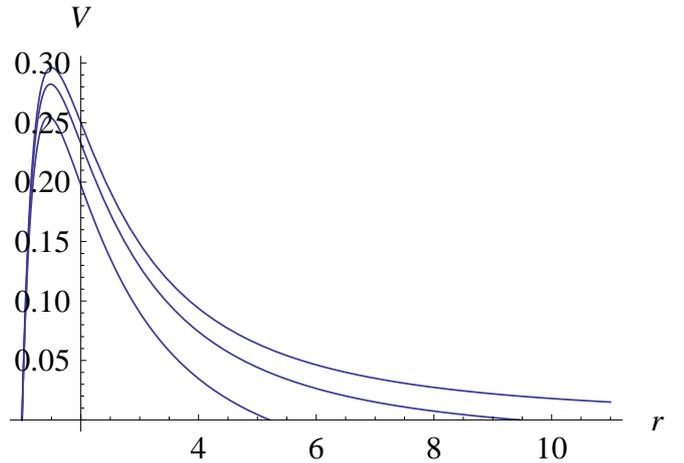}}
\caption{The effective potential for an electromagnetic perturbations in the background of the Schwarzschild-de Sitter black hole: $r_{0}=1$, $\ell=1$,  $\Lambda = 0$ (top), $\Lambda=0.003$, $\Lambda =0.09$ (bottom). The potentials have the ``canonical'' WKB form with a single maximum and monotonic decay at the boundaries. }\label{fig1}
\end{figure}

At high $\ell$, once the effective potential has the form of the potential barrier, falling off at the event and de Sitter horizons, the WKB formula found in \cite{Schutz:1985zz} can be applied for finding quasinormal modes:
\begin{equation}
\frac{Q_0(r_0)}{\sqrt{2Q_0^{''}(r_0)}}=i(n+1/2). \label{wkb}
\end{equation}
Here, the function $Q = \omega^2 - V$ and its second derivative $Q_0^{''}\equiv d^2Q_0/dr_*^2$ are evaluated at the extremum $r_{0}$.  It is believed that in the regime $\ell \rightarrow \infty$ the WKB formula (\ref{wkb}) is {\it exact}.
In the limit $\ell \gg n$ we have
\begin{equation}
Q_0\simeq \omega^2-f\frac{\ell^2}{r^2}.
\end{equation}
Taking the derivative of the above equation, one can observe that
\begin{equation}\label{extremum}
2 f(r_0)=r_0 f'(r_0),
\end{equation}
 The WKB formula (\ref{wkb}) then produce exactly what is expected from the null geodesics:
\begin{equation}\label{main}
\omega_{\rm QNM}=\ell \sqrt{\frac{f_{0}}{r_0^2}}
-i\frac{(n+1/2)}{\sqrt{2}}
\sqrt{-\frac{r_0^2}{f_{0}}\,\left (\frac{d^2}{dr_*^2}\frac{f}{r^2}\right )_{r_0}}.
\end{equation}

%\section{Counterexample: asymptotically de Sitter black holes}

An example we will consider here, the Schwarzschild-de Sitter black hole, is very well studied both as to the geodesic motion \cite{Stuchlik1,Stuchlik2} and quasinormal modes
\cite{Zhidenko:2003wq,Molina:2003dc,Konoplya:2004uk,Cardoso:2017soq,Konoplya:2022xid,Konoplya:2022kld,Churilova:2021nnc}. The radii of circular null geodesics are known to be independent on the cosmological constant \cite{Stuchlik1,Stuchlik2} and well described by the above formula.

The Schwarzschild-de-Sitter black hole is described by the metric function
\begin{eqnarray}\label{SdS-metric}%
f(r) =g(r) &=& 1 - \frac{2M}{r} - \frac{\Lambda r^2}{3}
%\JCAPstyle{=}
\\\nonumber&=&
\Lambda\frac{(r_{ch}-r)(r-r_0)(r+r_0+r_{ch})}{r}.
\end{eqnarray}
Here the black hole mass
$$M=\frac{r_{ch} r_0 (r_{ch} + r_0)}{2 (r_{ch}^2 + r_{ch} r_0 + r_0^2)},$$
and the cosmological constant
$$\Lambda=\frac{3}{r_{ch}^2 + r_{ch} r_0 + r_0^2},$$
can be expressed in terms of the radii of the event and cosmological horizons $r_0$ and $r_{ch}$.

The effective potential is known to have canonical form for both test boson and gravitational fields (see, for example, fig. \ref{fig1} for an electromagnetic field). Thus, all the arguments above are automatically applied and there must be $\ell \gg n$ quasinormal modes which are described by the parameters of the null geodesics. Indeed, it is well-known that the appropriate quasinormal modes are given by the following relation \cite{Zhidenko:2003wq}:
\begin{equation}
\omega = \frac{\sqrt{1- 9 M^2 \Lambda}}{3 \sqrt{3} M}\left(\ell + \frac{1}{2} + i \left(n +\frac{1}{2}\right)\right)+\mathcal{O}\left(\frac{1}{\ell}\right).
\end{equation}
The above formula is the one predicted by the null geodesics (\ref{QNM}). These quasinormal modes are Schwarzschild frequencies which are corrected by the cosmological constant, so that when $\Lambda \rightarrow 0$, they approach their Schwarzschild values \cite{Mashhoon:1982im}.

However, as was shown for the scalar field perturbations in  \cite{Cardoso:2017soq} and for gravitational ones in \cite{Konoplya:2022xid} there is another branch of quasinormal modes which are purely imaginary, that is, non-oscillatory exponentially decaying modes. When the radius of the black hole is much smaller than the radius of the cosmological horizon, there modes obey the following {\it universal law} \cite{Konoplya:2022kld}, which is valid not only for the Schwarzschild-de Sitter black hole, but for any spherically symmetric black hole independently on an underlying metric gravitational theory:
\begin{equation}\label{BHmodes}
  \omega_n = \omega_{n}^{(dS)}\left(1-\frac{M}{r_{ch}}+\Order{\frac{M}{r_{ch}}}^2\right).
\end{equation}
Here $\omega_{n}^{(dS)}$ are quasinormal modes of the empty de Sitter spacetime ~\cite{Lopez-Ortega:2006tjo,Lopez-Ortega:2006aal},
\begin{equation}\label{dSmodes}
\omega_{n}^{(dS)}r_{ch} = -\imo (\ell + n+1-\delta_{s0}\delta_{n0}),
\end{equation}
where $n=0,1,2,\ldots$ is the overtone number and $\ell=s,s+1,\ldots$ is the multiple number. Thus, when $r_{0}/r_{ch} \to 0$, the purely imaginary quasinormal frequencies go over into the modes of the empty de Sitter spacetime.

These purely imaginary quasinormal modes of asymptotically de Sitter black holes cannot be obtained with the help of the WKB formula even when $\ell \gg 1$.
The WKB formula is based on the WKB series expansions near the event and cosmological horizons which are matched with the Taylor expansion used in the middle region between the ``turning points'' $V - \omega^2 =0$ in the complex plain. For the Taylor expansion to be accurate, the turning points must be close. The real oscillation frequency is essentially determined by the height of the potential barrier in the limit $\ell \gg 1$, while the damping rate (given by the imaginary part of $\omega$) depends on the overtone number and second derivative of the effective potential. The WKB formula cannot be applied (see, for example, \cite{Iyer:1986np,Kokkotas:1988fm,Konoplya:2003ii,Matyjasek:2017psv,Konoplya:2019hlu}) once
\begin{equation}
\frac{Q_0(r_0)}{\sqrt{2Q_0^{''}(r_0)}} \gg 1,
\end{equation}
or, alternatively,
\begin{equation}
Re (\omega) \ll Im (\omega).
\end{equation}
Consequently, this branch of the purely imaginary quasinormal modes cannot be reproduced with the WKB method in principle, despite the effective potential looks ``WKB good''
(fig.\ref{fig1}). Moreover, when the cosmological constant is small enough, these purely imaginary modes have the smallest damping rate and dominate in the spectrum over the Schwarzschildian branch. Thus, even though the basic duality relation (\ref{QNM}) is formally fulfilled for this case, {\it $n$ is not the overtone number} for asymptotically de Sitter black holes, but numbers only the Schwarzschildian branch of the spectrum. We can remember that the eikonal approximation is usually used in the context of the limit of geometrical optics, that is, when the real oscillation frequency is very large $Re (\omega) \gg 1$. This is frequently replaced with the regime $\ell \gg 1$, which, as we see here, is not always identical. Thus, we can say that the duality indeed holds for the eikonal regime of quasinormal modes, but it may not reproduce the whole quasinormal spectrum, so that $n$ is not necessarily the overtone number. The same effect will evidently take place for rotating and higher dimensional black holes, because quasinormal modes must, anyway, approach the limit of the empty de Sitter spacetime.

The described here phenomenon is related to a kind of strong dependence of the quasinormal spectrum upon the tiny deformations of the geometry near the cosmological horizon, because even a very small $\Lambda$ leads to the appearance of purely imaginary modes which dominate at late times and cannot be reproduced via the duality.
In this connection, an interesting observation was made in  \cite{Price:2017cjr,Khanna:2016yow}: When the event horizon is replaced by the reflecting surface \cite{Price:2017cjr} or a wormhole throat \cite{Khanna:2016yow}, the correspondence (\ref{QNM}) is not observed. This may mean that the duality, which relies upon the WKB formula, may be broken once the geometry is deformed near the boundaries (either cosmological and/or event horizon) in such a way, that essentially non-WKB frequencies appear in the spectrum.

Taking all the above into account, we conclude that one should carefully interpret the correspondence between the eikonal quasinormal modes and characteristics of the null geodesics even in cases when the effective potential is ``WKB well-behaved'' and the formula (\ref{QNM}) can be formally applied.

\acknowledgments{The author thanks support of Czech Science Foundation (GAČR) grant 19-03950S.}

\end{document}